\documentclass{article}

\usepackage{arxiv}

\usepackage[utf8]{inputenc} 
\usepackage[T1]{fontenc}    
\usepackage{hyperref}       
\usepackage{url}            
\usepackage{booktabs}       
\usepackage{amsfonts}       
\usepackage{nicefrac}       
\usepackage{microtype}      
\usepackage{lipsum}
\usepackage{graphicx}
\graphicspath{ {./images/} }
\usepackage{multirow}

\title{A two-stage model leveraging friendship network for community evolution prediction in interactive networks}

\author{
 Yanmei Hu \\
  College of Computer and Cyber Security \\
  Chengdu University of Technology \\
  \texttt{huyanmei@cdut.edu.cn} \\
   \And
  Yihang Wu \\
  College of Computer and Cyber Security \\
  Chengdu University of Technology \\
  \texttt{wuyihang@stu.cdut.edu.cn} \\
  \And
 Biao Cai\\
  College of Computer and Cyber Security \\
  Chengdu University of Technology \\
  \texttt{caibiao@cdut.edu.cn} \\
}

\begin{document}
\maketitle
\begin{abstract}
Interactive networks representing user participation and interactions in specific ``events'' are highly dynamic, with communities reflecting collective behaviors that evolve over time. Predicting these community evolutions is crucial for forecasting the trajectory of the related “event”. Some models for community evolution prediction have been witnessed, but they primarily focused on coarse-grained evolution types (e.g., \textit{expand, dissolve, merge, split}
), often neglecting fine-grained evolution extents (e.g., the extent of community expansion). Furthermore, these models typically utilize only one network data (here is interactive network data) for dynamic community featurization, overlooking the more stable friendship network that represents the friendships between people to enrich community representations. To address these limitations, we propose a two-stage model that predicts both the type and extent of community evolution. Our model unifies multi-class classification for evolution type and regression for evolution extent within a single framework and fuses data from both interactive and friendship networks for a comprehensive community featurization. We also introduce a hybrid strategy to differentiate between evolution types that are difficult to distinguish. Experimental results on three datasets show the significant superiority of the proposed model over other models, confirming its efficacy in predicting community evolution in interactive networks.

\end{abstract}
\keywords{Community evolution prediction\and interactive networks \and classification \and regression}


\section{Introduction}
In real-world social networks, the emergence of communities is common, representing certain interaction patterns among various nodes within the network. Some of the social networks always keep changing, with communities in them evolving , and so are dynamic. For example, the interactive networks formed by users and their behaviors such as liking, forwarding and commenting on each other on social platforms. A community may undergo various evolutionary behaviors, including
 \textit{form} (the community is formed), \textit{expand}, \textit{shrink}, \textit{dissolve} (the community disappears), \textit{merge} (the community merges with other communities), and \textit{split} (the community splits into multiple communities) \cite{hu2016local}. The evolutions of communities in interactive networks are usually closely related to the occurrence of certain events, e.g., communities involved in emergencies tend to expand explosively in a short period of time. Therefore, predicting the evolutions of communities in the interactive social networks can not only foresee the changing trends of these networks, but also help identifying important events in advance.

\begin{figure*}
    \centering
    \includegraphics[width = \textwidth]{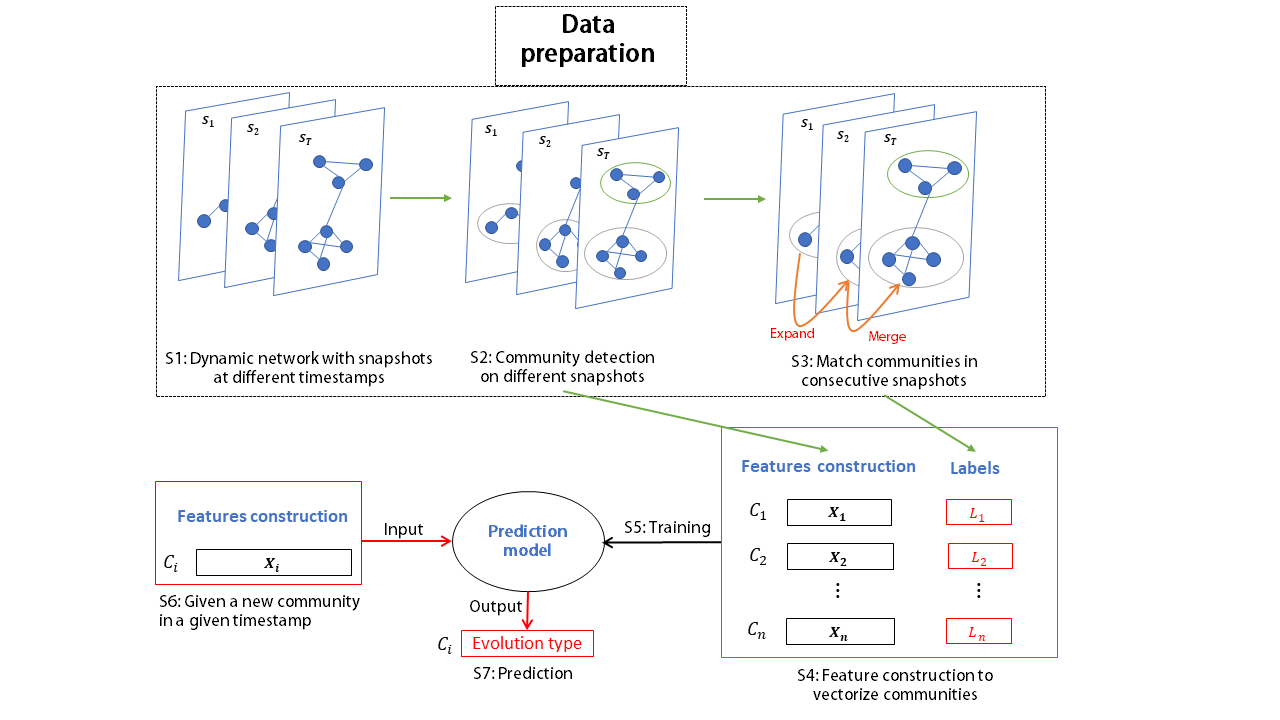}
    \caption{The general framework of community evolution prediction}
    \label{instance}
\end{figure*}

Many researchers approached the prediction of community evolution as a multi-classification problem, where a community corresponds to a sample with features generally extracted from the community's topological structure  and temoproal information, and the label of a community (i.e., the evolution type of the community) in the training set was obtained by matching communities in consecutive timestamps. See Fig.\ref{instance} for the framework of this approach. Under this framework, many methods were proposed. For example, Takaffoli et al. combined the attributes of influential members with topological structural features, including density, cohension, and coeifficient into a feature set which is used to vectorize the community, and several machine learning models such as SVM, Bagging, and BayesNet were used to train the multi-classification classifiers \cite{takaffoli2014community}. Makris et al. extracted subgraphs by random extraction from community and used Louvain algorithm \cite{que2015scalable} on subgraphs to obtain the features, and multi-class logistic regression was used to finish the multi-classification models \cite{makris2020distributed}. 
Wang et al. proposed constructing features by non-negative matrix factorization (NMF) on a hypergraph with nodes representing communities and edges representing connections between communities, and then employed logistic regression to classify the evolutions of communities \cite{wang2022multi}.

However, community evolution behavior encompasses both the type and extent of evolution. Understanding the extent of community evolution across various types is essential for deciphering the interactions of communities and enhancing the effectiveness of associated tasks. In addition, information about nodes in the other networks may facilitate the evolution prediction of their communities within the current network, as nodes typically belong to multiple networks. For example, a user in the platform of Sina Weibo belongs to the friendship network representing the follow and following relationship between users, and simultaneously, belongs to some interactive networks if this user participates in the corresponding “topics” (e.g., he/she commented or forwarded the “Henan covid-19” topic, and participate in the interactive network developed from this topic). It is easy to understand that, in this case, the position of this user in the friendship network would affect the evolution of its belonging to community in the interactive networks as this user's followers are the potential members of the interactive networks and may contribute to the evolution of its belonging to community in the interactive networks. Considering these two issues discussed above, we propose a two-stage model that leverages information from both the interactive and friendship networks, referred to as LTSModel, to predict both the type and extent of community evolution in interactive networks.
Specifically, the LTSModel integrates classification and regression into a single framework. Within this model, the classification component forecasts the type of community evolution, while the regression component predicts the extent of evolution for the given community type, as determined by the classification component. By sequentially optimizing the classification loss and the regression loss, both components of the proposed model are effectively trained. Furthermore, the LTSModel generates features for each community through a dual approach: characterizing the community's topological structure as well as temporal information and leveraging the positional properties of its members within the friendship network.
The main contributions of our work are as follows:

\textbf{1) A two-stage model that integrates classification and regression within one framework, enabling the prediction of both type and extent of community evolution.} To predict both type and extent of community evolution, which is ignored by previous works, we integrate classification and regression into one framework. The classification stage of the proposed model predicts the evolution type of a community from its feature vector; the regression stage concatenates the predicted evolution type and the feature vector as input to predict the the extent of this community evolution (i.e., the change in the number of nodes within the community). From the prediction results of the two-stage model, the community evolution can be reflected from two aspects, i.e., evolution type and extent.

\textbf{2) Leveraging information from the interactive network and friendship network to extract features for prediction.} When featurizing a community for subsequent evolution prediction, existing works have focused solely on the network containing the community, neglecting other networks to which the member nodes also belong. It is common for the same nodes to participate in multiple networks for various reasons, and a node's position in one network may significantly influence its neighborhood structure in others. Considering this, we extract features from both the interactive network and friendship network to represent community more expressively.

\textbf{3) Conducting experiments on real datasets to verify LTSModel.} We test the proposed model on two sets of real datasets. Experimental results show that LTSModel achieves much higher accuracy than the compared models on the type prediction of community evolution, and also yields low Mean Absolute Percentage Error (MAPE) on the extent prediction of community evolution, demonstrating its capability of predicting community evolution from both type and extent.

The organization of the subsequent content is as follows. Section 2 describes the related works about community evolution prediction. Section 3 defines the problem we focus on and the notations used in this paper. Section 4 presents the proposed model in details. Section 5 describes the used datasets and compare our model with other models. Section 6 is the final section and concludes the work.

\section{Related work}
In this section, we first describe the methods designed for community evolution prediction, and then present some other works related to our work.

Takaffoli et al. \cite{takaffoli2014community} proposed inputting changes in features representing the topological structure of communities into various traditional machine learning models, such as SVM and Bayesian classifiers, to predict community evolution. These features include community size, average closeness centrality, average degree centrality, and average eigenvector centrality. Additionally, the average closeness, degree, and eigenvector values of influential members were also considered as part of the feature set. Similar work was also observed in \cite{dakiche2021tailored,saganowski2019analysis}, but the classifiers were trained directly on the features, not the change of features.  
Chen et al. \cite{chen2022community} proposed integrating common topological features, including community size, average degree value, and community density, with latent features extracted through DeepWalk and spectral propagation techniques. A Random Forest model was then trained using the obtained features to predict community evolution.
Revelle et al. \cite{revelle2021group} designed a deep learning model called Group-Node Attention Network(GNAN) for community evolution prediction . GNAN takes the features representing the topological structure of the group and the degree information of group members as input, and employs a group-node attention mechanism and a fully connected layer to integrate the group features and member features.
Dakiche et al. \cite{dakiche2018sensitive} used the size, density, average betweenness values, and cohesion of the community as features, and applied logistics regression model as classifier .
Makris et al. \cite{makris2020distributed} proposed utilizing bagging approach to predict community evolution types. They first randomly extracted subgraphs from the network and trained a weak classifier based on multiple logistic regression on each subgraph, and finally assembled all the weak classifiers into a strong classifier used to complete the prediction. For each weak classifier, the training set was prepared by featurized the communities obtained by Louvain on the subgraph.
Wang et al. \cite{wang2022multi} took each community as a super node and constructed a hypergraph based on the super nodes and their connections, and extracted features for each community using NMF on the hypergraph. The classifier used to predict community evolution was trained using logistic regression. 
Man et al. \cite{man2019multi} proposed a chain-based classification algorithm for predicting community evolution. This algorithm constructs a chain-based classifier by cascading multiple base classifiers, which include a Markov chain model under the assumption that the evolution of a community adheres to a Markovian process. Within this chain-based classifier framework, the output from each preceding base classifier serves as part of the input for the subsequent base classifier, and state transition probability matrix were taken as features to represent each community.
Our work also extracts features to represent each community and constructs a predictive model using these feature vectors. Unlike existing approaches, our model uniquely integrates classification and regression within a single framework, enabling it to forecast both the type and extent of community evolution. In addition, the feature extraction process encompasses not only the topological structure of the community but also the positional characteristics of members within associated network, with a particular emphasis on friendship network.

Meanwhile, there are some other researches related to our work. In order to enhance the efficiency of community evolution prediction, some researchers have made efforts analyzing feature importance to community evolution or improving community detection. 
Boujlaleb et al. \cite{boujlaleb2020feature} used filter-based feature selection techniques to obtain the feature subsets related to each community evolution type, and considered the features with high frequency in the selected feature subsets as important features. Cheng et al. \cite{cheng2018novel} proposed searching the largest cliques within a network starting from the most active nodes, and each subgraph formed by these connected cliques was considered a community. In this approach, each clique was treated as a super node, with an edge between two super nodes indicating that the corresponding cliques share common nodes. Two cliques were considered connected if the associated super nodes were reachable from one another within the super graph. Yuan et al. \cite{yuan2022community} employed graph embedding techniques to forecast the evolution trends of user groups' preference for product tags in e-commerce. Initially, they constructed a bipartite graph comprising two types of nodes: user groups and tags. An edge between a user group node and a tag node indicated that at least one user within the group had purchased a product associated with that tag. Subsequently, they developed a hyper graph based on the bipartite graph, where each tag was represented as a node, and connections were made between tags that were linked by the same user group. GraphSage was applied to the bipartite graph, while a Graph Convolutional Network (GCN) was utilized on the hyper graph to generate two distinct embeddings for each tag. These embeddings were concatenated to form the final embedding representation for each tag. Furthermore, the historical embeddings of each tag and the embedding of a user group were input into a Gated Recurrent Unit (GRU)-based model to predict the user group's preference for this tag. In this study, we utilize the current state of a community at a given timestamp to predict its evolution in the next timestamp, and leave the prediction based on a longer sequence of historical states for future research.

\section{Problem statement and notations}
\textbf{Problem statement}. Given a community at timestamp $t$ in an interactive network, there are two tasks we aim to address. The first one is that which type of evolution this community tend to undergo in the next timestamp and the second one is to estimate the change in the number of nodes within this community (i.e., the extent of evolution) under the evolution type obtained in the first task. According to the literature \cite{hu2016local}, there are seven critical types of community evolution, which are \textit{form, continue, dissolve, expand, shrink, merge, split}. 
\begin{itemize} 
\item \textit{Form} refers to the formation of a new community as nodes in the interactive network organize into a new group over time. 
    \item \textit{Continue} and \textit{dissolve} refer to the continuation and dissolution of an existing community, respectively.
    \item      \textit{Expand} and \textit{shrink} refer to the increase and decrease in the number of members of an existing community, respectively.
    \item \textit{Merge} and \textit{split} refers to the combination of multiple communities into one and the division of one community into multiple communities, respectively.

\end{itemize}

Predicting the \textit{form} of a new community (whether a new community forms or not) entails examining all subsets of nodes within the interactive network, which leads to an exponential increase in time complexity. Consequently, this study does not include the prediction of \textit{form}; instead, we defer this aspect to future research. For a community, predicting its evolution type is inherently a multi-classification problem. Once the type of community evolution is confirmed, predicting the extent of its evolution can be approached as a regression problem. In this study, we introduce a two-stage model designed to address both of these two problems within a single, integrated framework.

\textbf{Notations}. Let $G_{int} = \{G_{int}^1,G_{int}^2,\ldots,G_{int}^t,\ldots,G_{int}^T\}$ represents the interactive network, where $G_{int}^t = \{V_{int}^t, E_{int}^t\}$ denotes the network snapshot at timestamp $t$, and $ V_{int}^t$ is the set of users in the interactive network up to timestamp $t$, and $E_{int}^t$ the set of edges induced by these users, indicating their interactions. The set of neighbors of user $v$ at timestamp $t$ is denoted as $N_v^t$, and $d_v^t$ denotes the degree of $v$ at timestamp $t$ (i.e., $d_v^t = |N_v^t|$). $C_i^t$ refers to a community in $G_{int}^t$, and $extd_v$ is the sum of degrees of $v$ and its neighbors. $durt(v)$ counts the number of snapshots in which $v$ appears. The friendship network is denoted by $G_{fri} = \{V, E_{fri}\}$, where $V$ is the set of all potential users, and $E_{fri}$ represents the friendships between them. $N_v^f$ is the set of neighbors of $v$ in $G_{fri}$, and $d_{max}^{fri}$ is the maximum degree in friendship network. For friendship network $G_{fri}$, there are also communities with each representing a social circle and denoted as $C_i^{fri}$. $C^{fri}(v)$ indicates the social circle to which $v$ belongs, and $Cid^{fri}(v)$ is the id of that social circle. Finally, $m$ denotes the number of labels (i.e., evolution types) we aim to predict. Notably that when no specific timestamp is mentioned, the sub-index $t$ for the notations related to the interactive network will be omitted in the subsequent text.
  
\section{The proposed LTSModel} \label{LSTModel}
\begin{figure*}
    \centering
    \includegraphics[width = \textwidth]{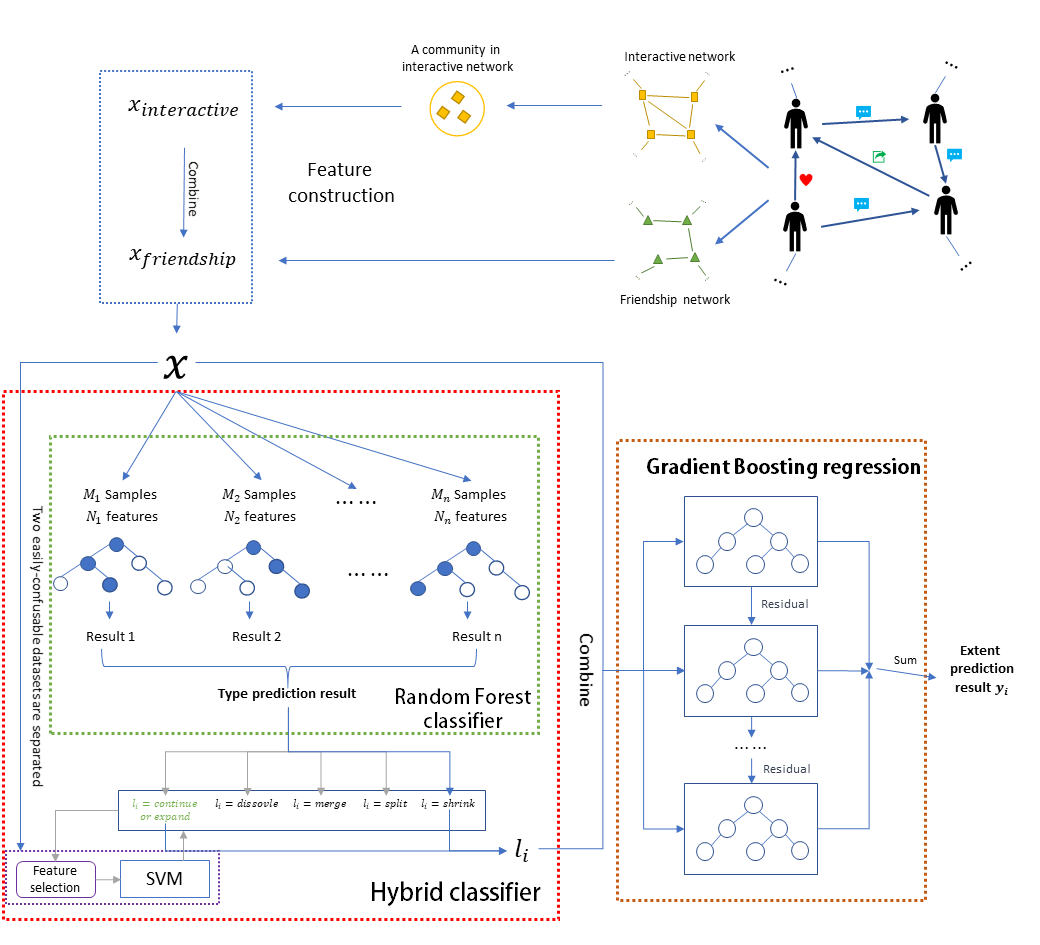}
    \caption{The framework of LTSModel, where the regressor is appended to the multi-classifier.
}
    \label{framework}
\end{figure*}

\subsection{The framework of LTSModel} \label{frameworkLSTM}
The framework of LTSModel is shown in Fig. \ref{framework}. There are three core components, which are feature construction, type prediction and extent prediction of community evolution. The feature construction component extracts features from two primary sources: firstly, the topological structure and temporal information of the community within its interactive network, and secondly, the positional properties of its members within the friendship network. These features are then leveraged to expressively represent the community. See the the upper part of Fig. \ref{framework} for the illustration of this component. The component of type prediction of community evolution is a hybrid multi-classification model, which is fed the feature vector obtained from “feature construction”, and combines random forest classifier and SVM to identify the type of community evolution. See the part enclosed by red rectangle in Fig. \ref{instance} for the illustration of this component. Notably, it is experimentally found that \textit{continue} and \textit{expand} are hardly separated from each other using general classification strategies, thus a feature selection algorithm from \cite{hu2023differential} is introduced to filter out noisy and redundant features. The extent prediction component is a regression model, which is appended to the classifier. It takes as input the feature vector from ``Feature Construction'' and the evolution type predicted by the multi-classification model, and employs a gradient boosting regressor to forecast the extent of community evolution. See the part enclosed by orange rectangle for the illustration of this component. Next, we first describe the three core components in details, and then briefly describe the train and inference process.

\subsection{Feature construction}\label{Feature-cosntruction}

Compared to interactive networks, friendship network exhibit greater stability. Potential users in interactive networks are typically from friendship networks, as each individual can be represented as a node in the friendship network—even those without friends, who can be considered isolated nodes. For a community in an interactive network, the positional properties of its members within the friendship network are likely to influence the evolution of this community; for instance, a node with more neighbors in the friendship network is more likely to attract additional participants to the interactive network. Therefore, to expressively featurize a community in the interactive network, we leverage information from both interactive and friendship networks. Specifically, for a community $C_i^t$ at timestamp $t$ in the interactive network, we employ 9 metrics to describe its topological structure and 2 metrics for its temporal information. Additionally, we use 4 metrics to measure the positional properties of its members in the friendship network, yielding a total of 15 features that characterize the community. These features and their definitions are presented in Table \ref{features}. To characterize the topological structure of $C_i^t$ in the interactive network, we consider its size, the edges induced by its members, the edges between its members, and the one-hop neighbors of its members, yielding 9 features. Specifically, $Nei(C_i^t)$, $Rei(C_i^t)$ and $Density(C_i^t)$ are derived from edges between members, reflecting $C_i^t$'s compactness. The features $d(C_i^t)$ and $Rd(C_i^t)$ account for all edges induced by its members, while $Extd(C_i^t)$ accounts for edges induced by members and their one-hop neighbors, reflecting $C_i^t$'s influence on non-members from some extent. The features $Reo(C_i^t)$ and $Cond(C_i^t)$ emphasize external connections, highlighting $C_i^t$'s connectivity with the remained network. Additionally, $Active(C_i^t)$ measures the compactness of $C_i^t$'s members at the previous timestamp, and $Persist(C_i^t)$ gauges the active duration of its members up to timestamp $t$.

To further describe the positional properties of $C_i^t$'s members in the friendship network, we introduce 4 metrics: the number of social circles containing the members ($Nsc(C_i^t)$), the sum of degrees of the members in the friendship network ($Sd(C_i^t)$), the number of social circles containing the neighbors of the members ($Sbdg(C_i^t)$), and the density of the members' social circles ($Density\_circile(C_i^t)$). These metrics provide insights into the social integration and influence of $C_i^t$'s members within the friendship network.

\begin{table*}
\centering
\caption{Features used to represent a community at timestamp t (denoted as $C_i^t$)}
\label{features}
\resizebox{\textwidth}{!}{%
\begin{tabular}{clclc}
\hline
\multicolumn{5}{c}{\textbf{Features from interactive network}} \\ \hline
\multicolumn{1}{c|}{\multirow{5}{*}{Structural}} &
  Community size &
  $|C_i^t|$ &
  \#Intra-edges &
  $Nei(C_i^t) =|\{ (v,u) \in G^{int}_t|u,v \in C_i^t \}|$ \\
\multicolumn{1}{c|}{} &
  Community degree &
  $d(C_i^t)=\sum_{v \in G^t_i}d_v$ &
  Ratio of intra-edges &
  $Rei(C_i^t) = \frac{Nei(C_i^t)}{|C_i^t|}$\\
\multicolumn{1}{c|}{} &
  Ratio of inter-edges &
  $Reo(C_i^t)= \frac{d(C_i^t)-Nei(C^t_i)}{|C_i^t|}$ &
  Ratio of degree &
  $Rd(C_i^t)= \frac{d(C_i^t)}{|C^t_i|}$ \\
\multicolumn{1}{c|}{} &
  Community density &
  $Density(C^t_i) = \frac{Nei(C^t_i)}{|C^t_i|(|C^t_i|-1)}$ &
  Community extended degree &
  $Extd(C_i^t)=\sum_{v \in C_i^t}extd_v$ \\
\multicolumn{1}{c|}{} &
  Conductance &
  $Cond(C^t_i)=\frac{d(C^t_i)-Nei(C^t_i)}{d(C^t_i)}$ &
  \multicolumn{1}{c}{} &
   \\ \hline
\multicolumn{1}{c|}{Temporal} &
  Activeness &
  $Active(C^t_i) = \frac{\left |\{ (u,v)\in G^{int}_{t-1}|u,v \in C^t_i \}\right |}{|C^t_i|}$ &
  Persistence &
  $Persist(C^t_i)=\frac{\sum_{v\in C^t_i}durt(v)}{|C^t_i|}$\\ \hline
\multicolumn{5}{c}{\textbf{Features from friendship network}} \\ \hline
\multicolumn{1}{c|}{\multirow{2}{*}{Structural}} &
  \#Social circles &
  $Nsc(C^t_i)= \left |\{ C_i^{fri}|\exists v \in C^t_i \&\& v \in C_i^{fri}\}\right|$ &
  Social degree &
  $Sd(C^t_i)=\frac{\sum_{v\in C^t_i}d_v^{fri}}{d_{max}^{fri}}$ \\
\multicolumn{1}{c|}{} &
  Social bridge &
  $Sbdg(C^t_i)= |\bigcup_{v\in C^t_i}{Cid^{fri}}(u)| u \in N_v^{fri} \&\& u \notin C^t_i \}|$&
  Density of social circles &
  $Density\_circle(C^t_i)= \sum_{v\in C^t_i} Density(C^{fri}(v))$ \\ \hline
\end{tabular}%
}
\end{table*}

\subsection{Type prediction of community evolution} \label{type-prediction}

A variety of multi-class classification models are candidates for the type prediction of community evolution, such as Decision Trees (DTs), K-Nearest Neighbors (KNN), and Neural Networks (NNs). Here we employ a Random Forest (RF) model, which constitutes an ensemble of multiple DTs, for the task at hand. This choice is motivated by superior expressiveness of RF compared to other traditional models, as well as its reduced requirement for an extensive feature set. In our specific case, the feature set is not overly extensive, rendering deep learning approaches less appropriate for our LTSModel. Consequently, the capabilities of RF model align more closely with the needs of our research, which is also validated by the experimental results as detailed in Section .\ref{type-results}, where RF performs much better than other models.

Within the RF model, each DT is constructed recursively and each node of the DT is generated as following. For a feature $A$ from the feature set, the training set, denoted as $T$, is divided into subsets based on the values of samples on $A$. Supposing the criterion for division is that: if a sample $x_i$ meets the condition $a_{j-1} <= x_i(A) < a_j$ , it is allocated to the $j$-th subset, denoted as $T_j$; then the Gini index for this division is defined as:
\begin{equation} 
G(T, A, a_1,…,a_j, …, a_k) = \sum_{i=1}^{k} {\frac{|T_i|}{|T|}(Gini(T_i))},\label{eq1} 
\end{equation}
\begin{equation}
    Gini(T_i) = 1 - \sum_{j=1}^{j=m} {p_{i,j}^2}, \label{eq2}
\end{equation}
\begin{equation}
    p_{i,j} = \frac{|\{x_i|l_i = j\}|}{|T_i|}.  \label{eq3}
\end{equation}
In Eq. \ref{eq3}, $p_{ij}$ is the ratio of samples with label $j$ in $T_i$. According to Eq. \ref{eq1}, the Gini index values of each feature and the possible division points are determined. The feature with the division that minimizes the Gini index value is selected to split this node into $k$ subtrees, with each subtree corresponding to a subset of the training set. This recursive process continues until a stopping criterion is reached, such as reaching the maximum tree or exhausting the available features, resulting in a fully-grown decision tree. Notably, features used in parent nodes are not reused for splitting at the current node. Additionally, to reduce overfitting, each DT is constructed on a subset of features selected through bootstrap sampling, rather than the entire feature set. The RF model is assembled once all constituent DTs have been constructed, and the output is derived from the voting result of the predictions made by all the DTs within it. The generation of Decision Trees (DTs) within the Random Forest (RF) model is halted when the available features are exhausted.

Furthermore, due to the difficulty in distinguishing between \textit{continue} and \textit{expand} evolution types, we employ a hybrid strategy for their classification. Initially, both types are treated as a single category during the construction of the RF model. Subsequently, we utilize a feature selection algorithm from \cite{hu2023differential} to identify the most informative features related to the evolution type. Thereafter, a SVM is trained on these selected features to distinguish \textit{continue} and \textit{expand}. We opt for the feature selection algorithm from \cite{hu2023differential} because the majority of communities that have experienced \textit{continue} and \textit{expand} exhibit very similar values across many features. This algorithm is capable of eliminating redundant and non-informative features, thereby enhancing the distinguishability of these two evolution types.

\subsection{Extent prediction of community evolution}\label{extent-prediction}

Given a community represented by a feature vector, once its evolution type has been identified using the hybrid multi-class classification model presented in the previous subsection, a regression model is required to estimate the extent of its evolution regarding to the determined type. Here we apply Gradient Boosting Trees (GBTs), an ensemble regressor consisting of multiple regression trees.

Similar to RF, GBTs generates multiple regression trees iteratively based on the training data and combine them to enhance the predictive output. However, unlike RF, GBTs is a sequential model where each tree is focused on correcting the error of the previous one. The process begins with the average response variable (i.e., the evolution extent) across all samples (i.e., communities) in the training dataset to create the first tree. Subsequent trees are then generated by minimizing the negative gradient of the loss function, with each tree aiming to correct the error of its predecessor. Ultimately, the model converges on the true value through this sequence of trees. The output of GBTs with $k$ trees is formulated as follows:

\begin{equation}
    F_k(x) = F_{k-1}(x) + \gamma_k \times h_k(x)
\end{equation}
where $ F_{k-1}(x)$ is the output of the GBTs with the first $k-1$ trees, which is equivalent to $F_k(x)$ in formulation, $h_k(x)$ is the output of the $k$-th tree (i.e., the fitted error for the output of the $(k-1)$-th tree), and $\gamma_k$ is the learning rate associate with the $k$-th tree. $F_0(x)$ and $h_k(x$) are defined as follows:
\begin{equation}
    F_0(x) = \frac{\sum_{i=1}^{|T|}{y_i}}{|T|},
    \label{eq_mean}
\end{equation}
\begin{equation}
    h_k(x) \approx y - F_{k-1}(x).
    \label{eq_h}
\end{equation}
Eq. \ref{eq_mean} means that the first tree has only one node with the value set to the mean of the target values of all samples. Eq. \ref{eq_h} means that each subsequent regression tree tries to fit the residual of its predecessor.
For each tree within the GBT, its generation process is analogous to that in RF:
It begins with the first node corresponding to the training set, then calculates the Mean Square Errors (MSEs) assuming splitting the node according to each feature with possible division points to find the best splitting point, finally splits the node into subtrees with each subtree corresponding to a subset of the training set. This recursive process continues until the stopping criterion, e.g., reaching the maximum tree and exhausting the available features, is met.

\subsection{Training and inference}\label{ti}
\textbf{Training.} 
To train LTSModel, a set of communities with evolution behavior, denoted as $ T_{ori}= \{(C_i, l_i, y_i) | i=1,\ldots,n\}$, are required. These communities and their evolution behaviors can be obtained by observing the interactive networks. For example, given an interactive network $G_{int}$, we can collect communities from each observed snapshot. The evolution behavior of a community at timestamp $t$ can be ascertained by aligning it with communities in the subsequent timestamp $t+1$ (the matching method in \cite{hu2016local} is utilized to identify the evolution type, and the evolution extent is calculated as the difference between the numbers of nodes of the community and its aligned community at timestamp $(t+1)$). For each community $C_i$, we construct its feature vector $x_i$ as detailed in Section \ref{Feature-cosntruction}, forming the training set $T= \{(x_i, l_i, y_i) | i = 1,\ldots, n\}$. We then apply the pairs $(x_i, l_i)$ to the model presented in Section \ref{type-prediction} to train the multi-classification classifier. This classifier is a hybrid model integrating RF and SVM. Samples with $l_i$ labeled as \textit{continue} and \textit{expand} are grouped as one class for RF training; subsequently, they are used to train an SVM-based binary classifier post RF training. Thereafter, we merge each $x_i$ and $l_i$ into a new feature vector $x_{i}^{'}$, and apply the pairs $(x_{i}^{’}, y_i)$ to the model described in Section \ref{extent-prediction} to train the regression model.

\textbf{Inference.} Given a community $C_i$ in an interactive network, its evolution type and extent are predicted as follows: First, we construct its feature vector as detailed in Section \ref{Feature-cosntruction}. Second, we input this feature vector into the multi-class classifier to forecast its evolution type $l_i$ as illustrated in the red-bordered part in Fig. \ref{framework}. Third, we merge the predicted evolution type and the feature vector into a new feature vector, and input this new feature vector into the regressor to predict the evolution extent $y_i$, as illustrated in the orange-bordered part in Fig. \ref{framework}.

\section{Experiments}

We conduct two experiments to evaluate the proposed LTSModel. The first experiment involves comparing LTSModel with other methods, while the second focuses on assessing the critical components of LTSModel. Before presenting the results, we provide an overview of the datasets, the methods compared, and the experimental settings.

\subsection{Datasets  and baselines} \label{Secdatasets}

We collect 9 interactive networks from Sina Weibo, each stemming from a hot topic. For each hot topic, the interactive network is constructed by representing each participating user as a node and each interactive behavior (such as commenting or forwarding) as an edge. Nodes and edges are organized into distinct snapshots based on timestamps. For instance, if snapshots are taken every 30 minutes and user A comments on user B's post within the first 30 minutes, both users and the edge connecting them will be included in the first snapshot. The friendship network is derived from the follow relationships on Sina Weibo. Given that follow relationships between users are significantly more stable than the interactive behaviors regarding hot topics, we consider the friendship network to be static. Given the vast number of users and the limitations on data collection imposed by the platform, we obtain the friendship network for an interactive network by performing a breadth-first-search algorithm from each user involved in the corresponding hot topic, within a two-hop constraint. Statistical details of these networks are displayed in Table \ref{Sina}, where the ``\#Nodes" and ``\#Edges" columns list the number of nodes and edges in each interactive network or friendship network, the ``Periods" column indicates the duration of each interactive network. Since community structure is implicit within the interactive and friendship networks, we perform community detection to reveal them.  Specifically, we use LMC-APR \cite{hu2022exhaustive} and Label Propagation Algorithm (LPA) \cite{lpa2019} to identify communities in each snapshot of the interactive networks as well as the friendship network. For each community $C_i$ in the interactive networks, except those in the final snapshot, we convert it into a sample $(x_i, l_i, y_i)$ , with $x_i$ derived as described in Section \ref{Feature-cosntruction} and $l_i$ and $y_i$ obtained as outlined in the ``training" part of Section \ref{ti}. All samples are pooled into a single dataset, named SinaHotTopics , as they originate from the same platform.

Additionally, we also use two public datasets, SuperUser and StackOverflow, from the SNAP platform. The statistical information for these three datasets is presented in Table \ref{datasets}, where the columns ``\#Nodes" and ``\#Edges" indicate the numbers of nodes and edges, respectively, ``Periods" indicates the duration of each dataset, and ``\#Samples under LMC-APR" and ``\#Samples under LPA" indicate the numbers of samples using LMC-APR and LPA as community detection algorithm, respectively. It should be noted that friendship networks for the two public datasets are not available.

\begin{table*}
\centering
\caption{The information of interactive networks from Sina platform}
\label{Sina}
\resizebox{\textwidth}{!}{%
\begin{tabular}{c|ccc|cc}
\hline
 &
  \multicolumn{3}{c|}{Interactive networks} &
  \multicolumn{2}{c}{Friendship networks} \\ \hline
Datasets & \#Nodes & \#Edges & Periods(h) & \#Nodes & \#Edges \\ \hline
Xian\_covid & 3659 & 2951 & 3 & 83349125 & 11725228 \\ \hline
Xuzhou\_covid & 5749 & 3636 & 3 & 7725050 & 10860239 \\ \hline
Chengdu\_covid & 5126 & 7026 & 6 & 8126160 & 13675496\\ \hline
Beijing\_covid & 9996 & 8149 & 3 & 7726986 & 11726217\\ \hline
Xuzhou\_subway\_covid & 7355 & 5214 & 2 & 7105372 & 10548411 \\ \hline
Shanghai\_covid & 5288 & 6167 & 74 & 7418144 & 133919509\\ \hline
Express\_delivery\_covid & 29157 & 74167 & 2 & 7329007 & 11127427\\ \hline
Chongqing\_covid & 11815 & 17574 & 5 & 7954295 & 15198360\\ \hline
Omicron & 19969 & 49776 & 2 & 7121204 & 133017105\\ \hline
\end{tabular}%
}
\end{table*}

\begin{table*}
\centering
\caption{The statistical information of the three datasets.}
\label{datasets}
\resizebox{\textwidth}{!}{%
\begin{tabular}{c|ccccc}
\hline
Datasets &
  \#Nodes &
  \#Edges &
  Periods &
  \begin{tabular}[c]{@{}c@{}}\#Samples \\ under LMC-APR\end{tabular} &
  \begin{tabular}[c]{@{}c@{}}\#Smaples \\  under LPA\end{tabular} \\ \hline
SuperUser     & 194085     & 1443339& 2773(days)& 44148116 & 6707759 \\
StackOverFlow & 2601977    & 63497050& 2774(days)& 573519   & 221294  \\
SinaHotTopics & 1664-21450 & 1681-22995 & 2-74(hours)& 107506   & 42595   \\ \hline
\end{tabular}%
}
\end{table*}

Since there is lack of models that predict both evolution type and extent simultaneously, we test the two parts of LSTModel separately. Specifically, for the type prediction of community evolution, we compare LSTModel with the following models:

\textbf{Convolutional Neural Networks (CNNs) \cite{amerini2019social}}: Originally designed for image data, CNNs are now extensively applied to diverse datasets. CNNs utilize convolutional layers to identify patterns such as edges, pooling layers to reduce the spatial dimensions, and fully connected layers to compute class probabilities. 

\textbf{Multilayer Perceptron (MLP)}: A MLP classifies data by discerning patterns through a series of fully connected layers. Within each layer, neurons apply weights and biases to the input data, followed by the application of activation functions to the resulting outputs. The network's learning process involves mapping inputs to their respective classes by iteratively adjusting the weights during training, thereby minimizing the classification error. 

\textbf{CatBoost \cite{hancock2020catboost}}: CatBoost is a gradient boosting algorithm specifically optimized for handling categorical features. It efficiently processes categorical data, mitigates overfitting, and achieves high accuracy with minimal hyperparameter tuning, rendering it particularly suitable for tasks involving structured data. 

\textbf{XGBoost \cite{ramraj2016experimenting}}: XGBoost is a fast, scalable gradient boosting algorithm that optimizes tree-based models. Known for regularization, it reduces overfitting and improves accuracy, making it popular in structured data competitions and large datasets.

\textbf{LightGBM \cite{ke2017lightgbm}}: LightGBM is a gradient boosting framework that employs histogram-based learning for rapid training on large datasets. It is characterized by its memory efficiency, ability to handle categorical features, and its effectiveness with large-scale structured data.

\textbf{Support Vector Machine (SVM)}: SVM is a supervised learning algorithm used for classification. It finds the optimal hyperplane that best separates classes by maximizing the margin between them. SVM is effective in high-dimensional spaces and is versatile with different kernels for linear and non-linear data.

For the extent prediction of community evolution, we compare LSTModel with the following models:

\textbf{Random Forest for regression (RF(R))}:  RF(R) constructs an ensemble of decision trees trained on varied data samples, averaging their predictions to enhance accuracy. This approach mitigates overfitting and captures intricate data relationships. It demonstrates robustness to noise, effectively handles missing values, and performs well with structured data.

\textbf{Support Vector Regression (SVR) \cite{parbat2020python}}: SVR is an adaptation of SVM for regression tasks. Instead of finding a hyperplane to separate classes, SVR seeks a ``tube" around the hyperplane that best fits the data, minimizing error within a specified margin (epsilon). SVR is effective for high-dimensional data and supports various kernels for both linear and non-linear data.

\subsection{Experiment settings}

Each dataset was randomly split into training and testing sets following a ratio of 10:1 or 10:2 (StackOverFlow and SuperUser were split with ratio of 10:1, SinaHotTopics was 10:2 as this dataset was much smaller than the other two). To address the model training bias caused by the significant class imbalance among different evolution types, we employed oversampling on the underrepresented classes to achieve a balanced training dataset. Given the absence of friendship network data in the two public datasets, we set the corresponding elements in the feature vector to zero. All models were developed in Python 3.9 using PyCharm and executed on a PC with an Intel Core i9-10700K CPU and a Nvidia GTX 1650 GPU.

\subsection{The comparison results}

\subsubsection{The comparison results of evolution type prediction } \label{type-results}
The evaluation metrics for the prediction results of evolution type by different models include accuracy (Acc), Precision(P), weighted average Recall (R), and F1 score (F1). The results of different models using LMC-APR and LPA as community detection algorithm are presented in Tables \ref{clf_1} and \ref{clf_2}, respectively.

Table \ref{clf_1} indicates that LTSModel outperforms other models significantly in all cases when communities are revealed by LMC-APR. The improvements over other models range from 3.57\% to 481.40\% in ACC, 3.50\% to 3386.33\% in Precision, 3.57\% to 3386.33\% in Recall, and 3.64\% to 1935.92\% in F1. In addition, among the compared models, XGBoost performs best, followed closely by LightGBM, then CatBoost, with SVM yielding poor results and CNN and MLP being the worst. Table \ref{clf_2} shows similar trends with LPA as the community detection algorithm, demonstrating that LTSModel's superiority is consistent across different algorithms of community detection.

These findings suggest that LTSModel excels in predicting community evolution types in interactive networks, regardless of the community detection algorithm used.

\begin{table*}
\centering
\caption{The prediction results of evolution type by different models using LMC-APR as community detection algorithm}
\label{clf_1}
\resizebox{\textwidth}{!}{%
\begin{tabular}{ccccccccc}
\hline
Datasets& Metrics  & CNN    & MLP    & CatBoost & XGBoost & Lightgbm & SVM    & LTSModel\\ \hline
\multirow{4}{*}{SuperUser}         & Acc& 0.1667 & 0.1667 & 0.5462   & 0.8195  & 0.7808   & 0.3792 & \textbf{0.9692}            \\
                                   & P& 0.0278 & 0.0278 & 0.5612   & 0.8237  & 0.7871   & 0.4898 & \textbf{0.9692}            \\
                                   & R& 0.1667 & 0.1667 & 0.5462   & 0.8195  & 0.7808   & 0.3792 & \textbf{0.9692}            \\
                                   & F1& 0.0476 & 0.0476 & 0.5361   & 0.8174  & 0.7781   & 0.3287 & \textbf{0.9691}            \\ \hline
\multirow{4}{*}{StackOverFlow}     & Acc
& 0.1667 & 0.1667 & 0.4328   & 0.7425  & 0.7090   & 0.2112 & \textbf{0.9130}            \\
                                   & P
& 0.0278 & 0.0278 & 0.4772   & 0.7561  & 0.7239   & 0.2196 & \textbf{0.9225}            \\
                                   & R
& 0.1667 & 0.1667 & 0.4328   & 0.7425  & 0.7090   & 0.2112 & \textbf{0.9130}            \\
                                   & F1& 0.0476 & 0.0476 & 0.4304   & 0.7405  & 0.7046   & 0.1421 & \textbf{0.9112}            \\ \hline
\multirow{4}{*}{SinaHotTopics } & Acc
& 0.1667 & 0.1667 & 0.5417   & 0.9358  & 0.9200   & 0.3200 & \textbf{0.9692}            \\
                                   & P
& 0.0278 & 0.0278 & 0.5510   & 0.9364  & 0.9203   & 0.3262 & \textbf{0.9692}            \\
                                   & R
& 0.1667 & 0.1667 & 0.5417   & 0.9358  & 0.9200   & 0.3200 & \textbf{0.9692}            \\
                                   & F1& 0.0476 & 0.0476 & 0.5351   & 0.9350  & 0.9190   & 0.2826 & \textbf{0.9691}            \\ \hline
\end{tabular}%
}
\end{table*}

\begin{table*}
\centering
\caption{The prediction results of evolution type by different models using LPA as community detection algorithm.}
\label{clf_2}
\resizebox{\textwidth}{!}{%
\begin{tabular}{ccccccccc}
\hline
Datasets& Metrics  & CNN    & MLP    & CatBoost & XGBoost & Lightgbm & SVM    & LTSModel\\ \hline
\multirow{4}{*}{SuperUser}         & Acc
& 0.1673 & 0.1667 & 0.5483   & 0.8690  & 0.8432   & 0.2623 & \textbf{0.9660}            \\
                                   & P
& 0.2192 & 0.0278 & 0.5425   & 0.8677  & 0.8414   & 0.1693 & \textbf{0.9661}            \\
                                   & R
& 0.1673 & 0.1667 & 0.5483   & 0.8690  & 0.8432   & 0.2623 & \textbf{0.9660}            \\
                                   & F1& 0.0509 & 0.0476 & 0.5377   & 0.8672  & 0.8407   & 0.1631 & \textbf{0.9659}            \\ \hline
\multirow{4}{*}{StackOverFlow}     & Acc
& 0.1668 & 0.2312 & 0.4963   & 0.6782  & 0.6445   & 0.1808 & \textbf{0.9697}            \\
                                   & P
& 0.1944 & 0.1920 & 0.5126   & 0.6848  & 0.6506   & 0.1723 & \textbf{0.9702}            \\
                                   & R
& 0.1668 & 0.2312 & 0.4963   & 0.6782  & 0.6445   & 0.1808 & \textbf{0.9697}            \\
                                   & F1& 0.0480 & 0.1728 & 0.4745   & 0.6739  & 0.6383   & 0.0741 & \textbf{0.9696}            \\ \hline
\multirow{4}{*}{SinaHotTopics } & Acc
& 0.1667 & 0.2542 & 0.6308   & 0.9567  & 0.9550   & 0.2658 & \textbf{0.9625}            \\
                                   & P
& 0.0278 & 0.1162 & 0.6341   & 0.9576  & 0.9563   & 0.3574 & \textbf{0.9639}            \\
                                   & R
& 0.1667 & 0.2542 & 0.6308   & 0.9567  & 0.9550   & 0.2658 & \textbf{0.9625}            \\
                                   & F1& 0.0476 & 0.1417 & 0.6151   & 0.9557  & 0.9540   & 0.1894 & \textbf{0.9624}            \\ \hline
\end{tabular}%
}
\end{table*}

\subsubsection{The comparison results of evolution extent prediction} \label{extent-result}
MAPE is used as the evaluation metric for assessing the prediction of evolution extent, given the wide variation in evolution extents across different communities. Table \ref{reg} shows the MAPE values for different models. It can be seen that LTSModel and RF(R) yield significantly smaller MAPE values than SVR across all datasets, and LTSModel yields obviously smaller MAPE value than RF(R), irrespective of whether LMC-APR or LPA is used to reveal communities. Specifically, when communities are revealed by LMC-APR, LTSModel achieves a MAPE value of 0.9758 on SuperUser, reducing the MAPE by 98.7\% and 68.4\% compared to SVR and RF(R), respectively. On stackOverFlow, LTSModel achieves a MAPE value of 7.5286, with reductions of 90.7\% and 56.6\% compared to SVR and RF(R), respectively. On SinaHotTopics , LTSModel achieves a MAPE value of 1.8344, with reductions of 97.1\% and 68.1\% compared to SVR and RF(R), respectively. When communities are revealed by LPA, LTSModel achieves MAPE values of 2.5131, 7.1635 and 20.6453 on SuperUser, StackOverFlow and SinaHotTopics , respectively, with reductions of 96.6\% (83.7\%), 90.4\% (48.1\%), and 78.9 (36.6\%) compared to SVR (RF(R)).

\begin{figure}
    \centering
    \includegraphics[width=0.9\textwidth]{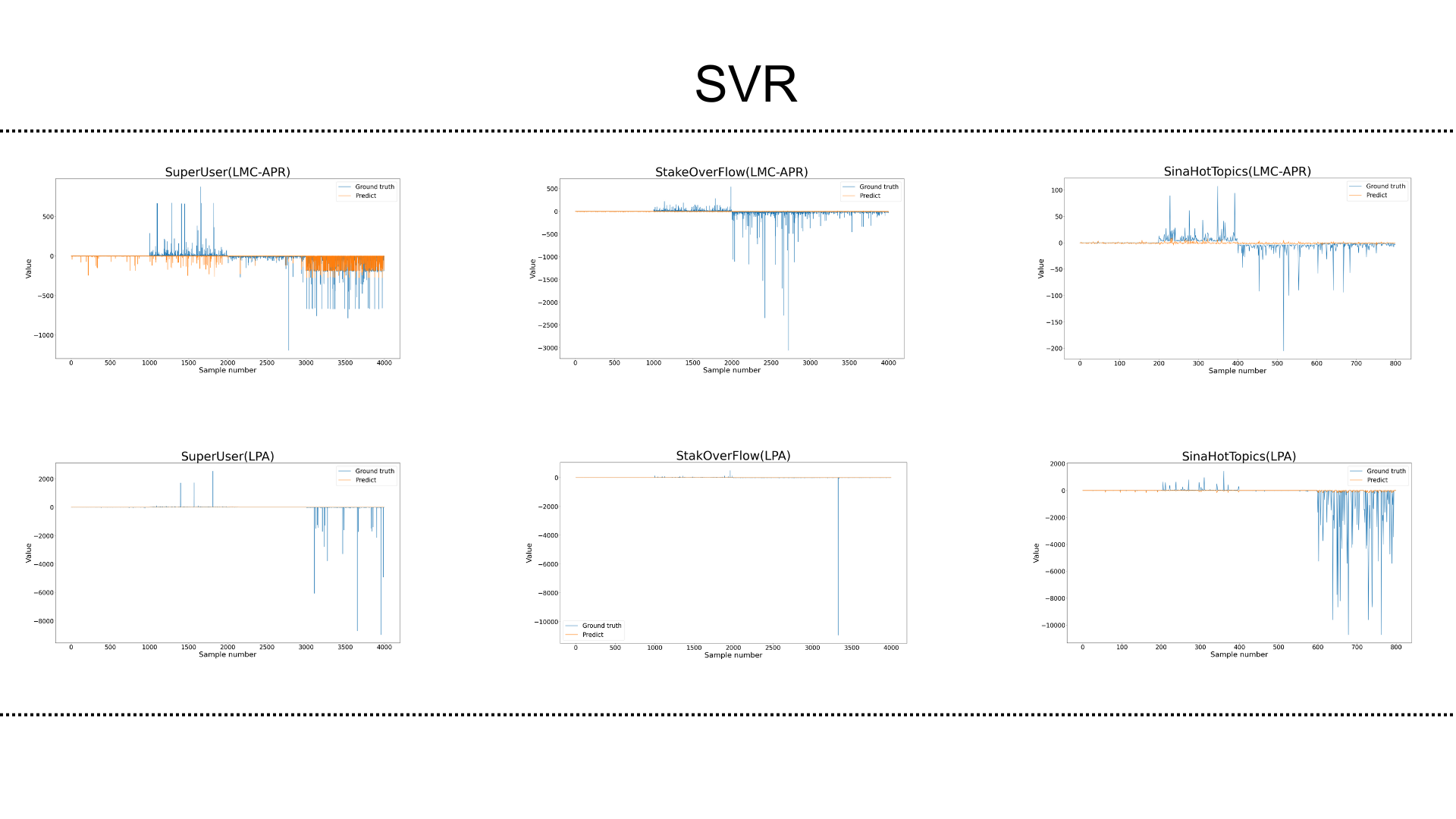}
    \caption{The visualization of evolution extent predicted by SVR}
    \label{fitness_1}
\end{figure}

\begin{figure}
    \centering
    \includegraphics[width=0.9\textwidth]{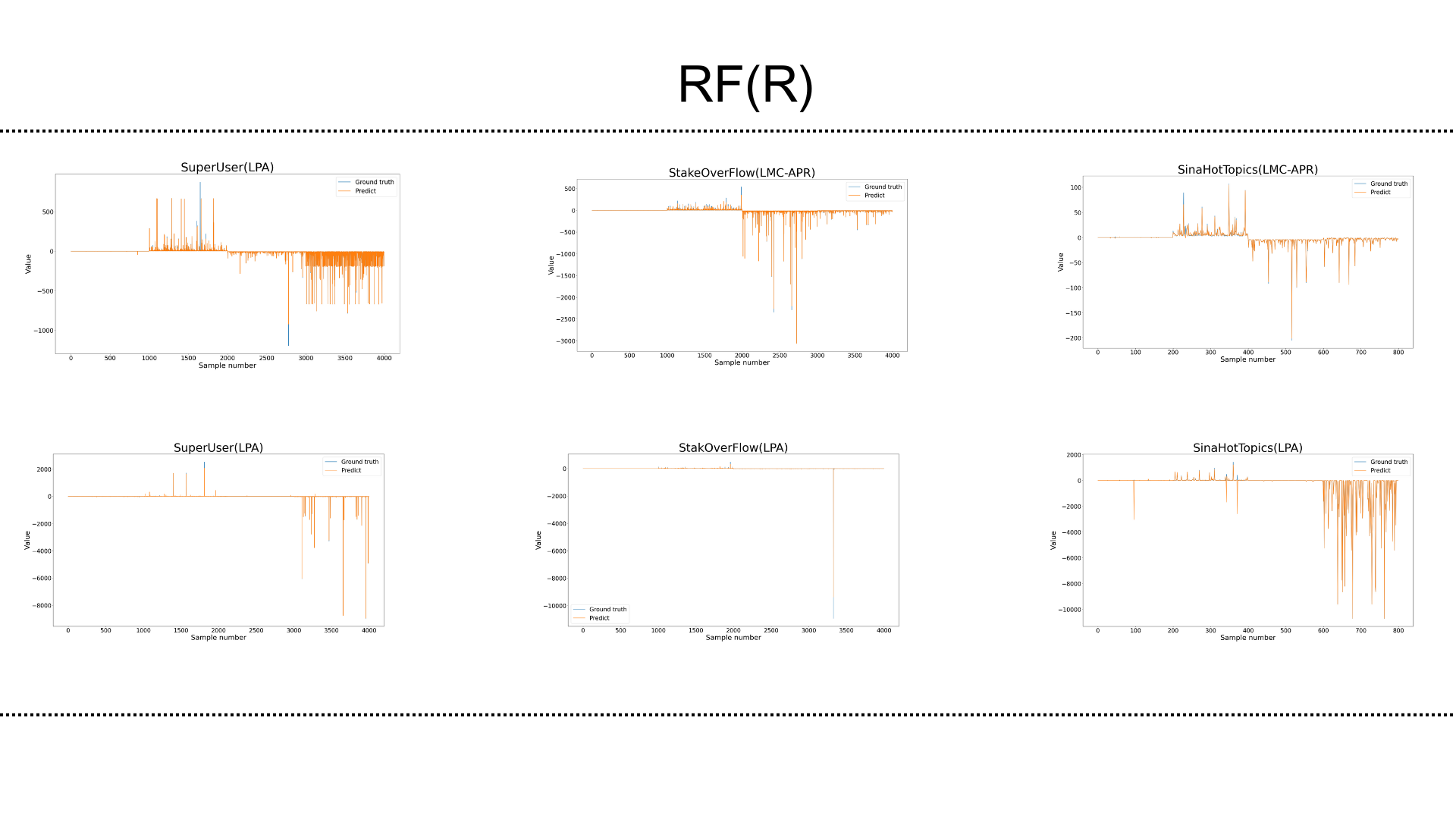}
    \caption{The visualization of evolution extent predicted by RF(R)}
    \label{fitness_2}
\end{figure}

\begin{figure}
    \centering
    \includegraphics[width=0.9\textwidth]{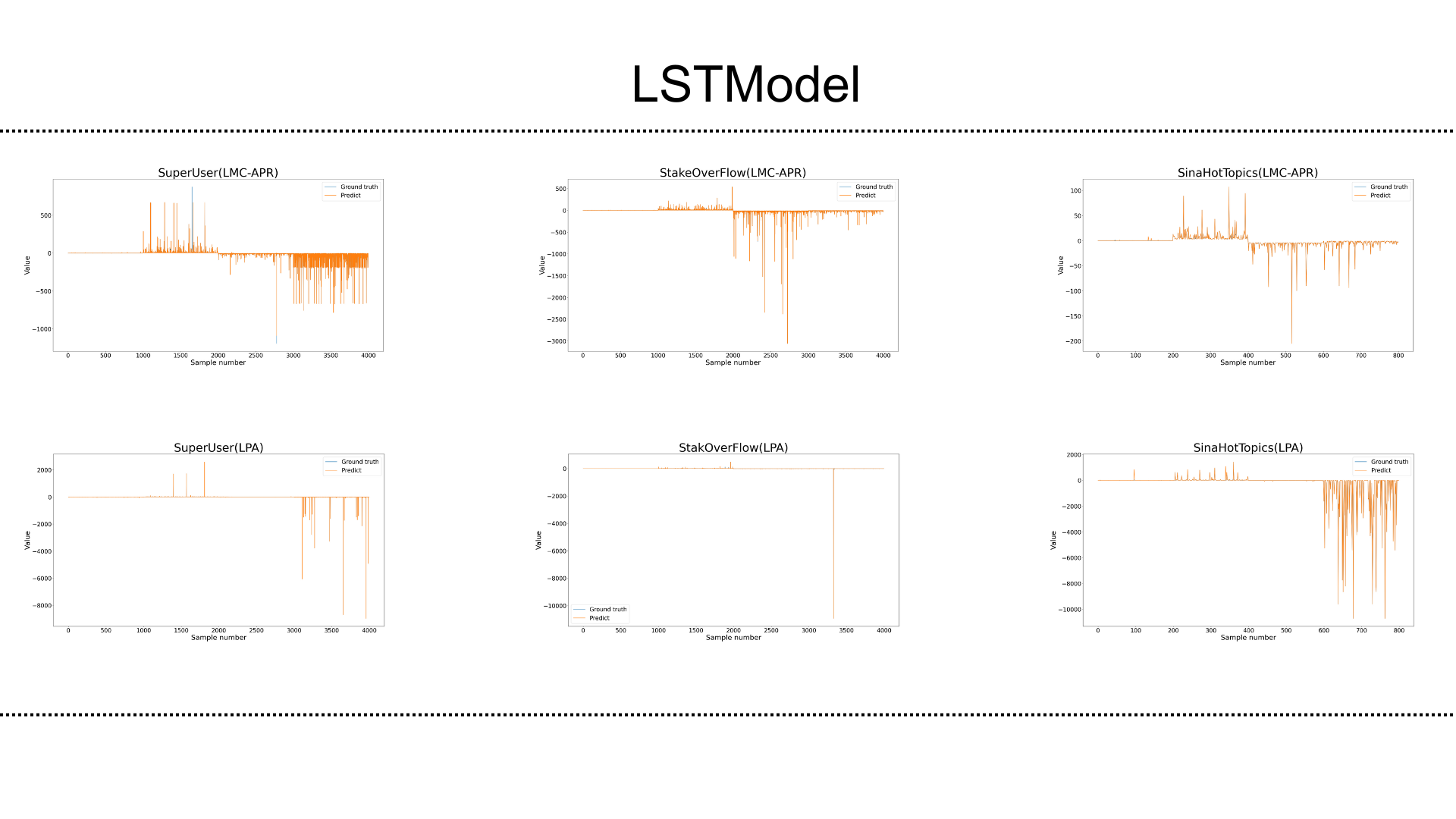}
    \caption{The visualization of evolution extent predicted by LTSModel}
    \label{fitness_3}
\end{figure}

Furthermore, the prediction performance of each model is visualized in Figs. \ref{fitness_1}, \ref{fitness_2}, and \ref{fitness_3}. Each chart corresponds to a dataset, with blue line representing the real evolution extents and orange line representing predicted evolution extents. The visualization aligns with the result in Table \ref{fitness_1}, showing that SVR generally fails to predict the evolution extent of most communities accurately, as indicated by larger discrepancies between the blue and orange lines and larger MAPE values. RF(R) achieves close prediction in many cases, with closer alignment between the blue and orange lines and smaller MAPE values. LTSModel successfully predicts the evolution extents in most cases, as evidenced by the high consistency between the blue and orange lines and the significantly smaller MAPE values. 

From the above discussions, it can be inferred that LTSModel outperforms the compared models in all cases, highlighting its superiority in predicting the evolution extent of communities.

\begin{table}
\centering
\caption{The prediction results of evolution extent by different models (MAPE)}
\label{reg}
\resizebox{\textwidth}{!}{%
\begin{tabular}{ccccc}
\hline
Datasets & \begin{tabular}[c]{@{}c@{}}Community detection \\ algorithm\end{tabular} & SVR & RF(R)& LTSModel\\ \hline
\multirow{2}{*}{SuperUser}     & LMC-APR & 75.1901 & 3.0874  & \textbf{0.9758}  \\
                               & LPA     & 74.6616 & 15.4193 & \textbf{2.5131}  \\ \hline
\multirow{2}{*}{StackOverFlow} & LMC-APR & 80.5474 & 17.3590 & \textbf{7.5286}  \\
                               & LPA     & 74.3615 & 13.7921 & \textbf{7.1635}  \\ \hline
\multirow{2}{*}{SinaHotTopics } & LMC-APR & 64.2025 & 5.7421  & \textbf{1.8344}  \\
                               & LPA     & 98.0633 & 32.5593 & \textbf{20.6453} \\ \hline
\end{tabular}%
}
\end{table}


\subsection{Ablation study}

In this experiment, we assess the impact of feature extraction from the friendship network and the hybrid classification strategy for distinguishing \textit{continue} and \textit{expand}. Table \ref{ablation_1} compares the results of LTSModel with TSModel, which is identical to LTSModel except that it extracts features solely from the interactive network (this ablation test was not done on SuperUser and StackOverFlow as the friendship network was absent in these two datasets). Table \ref{ablation_2} compares LTSModel with LTSModel (RF only), which lacks the hybrid classification strategy for \textit{continue} and \textit{expand}.

According to Table \ref{ablation_1}, LTSModel performs better than TSModel (without friendship network) in all cases. For SinaHotTopics,  when communities are revealed by LMC-APR, LTSModel enhances performance over TSModel by 1.04\% in Acc, 1.05\% in Precision, 1.04\% in Recall and 1.05\% in F1. When communities are revealed by LPA, LTSModel enhances performance over TSModel by 4.52\% in Acc, 4.44\% in Precision, 4.52\% in Recall and 4.56\% in F1. These improvements prove that friendship network bring useful information for community evolution prediction. However, the improvement seems marginal, which is probably because the friendship network is incomplete due to data limitation from the Sina Weibo platform. 

According to Table \ref{ablation_2}, LTSModel is superior to LTSModel (RF only) in all cases. Specifically, when communities are revealed by LMC-APR, LTSModel enhances performance over LTSModel (RF only) by 1.20\% to 3.19\% in Acc, 2.01\% to 3.15\% in Precision, 1.20\% to 3.19\% in Recall, and 1.38\% to 3.30\% in F1, with the most notable improvements observed on SinaHotTopics. When communities are revealed by LPA, LTSModel improves Acc by 0.52\% to 1.3\%, Precision by 0.55 \% to 4.88\%, Recall by 0.52\% to 4.94\%, and F1 by 0.60\% to 5.06\%, with the most notable improvements observed on StackOverFlow. These results underscore the substantial benefits of the hybrid classification strategy in enhancing LTSModel’s predictive capability.

\begin{table}
\centering
\caption{The prediction results of evolution type by LTSModel and TSModel (without friendship networks)}
\label{ablation_1}
\resizebox{0.5\textwidth}{!}{%
\begin{tabular}{c|c|ccc}
\hline
Datasets &
  \begin{tabular}[c]{@{}c@{}}Community \\ detection \\ algorithm\end{tabular} &
  Metrics &
  TSModel&
  LTSModel \\ \hline
\multirow{8}{*}{SinaHotTopics} &
  \multirow{4}{*}{LMC-APR} &
  Acc & 0.9592 & \textbf{0.9692} \\
 &                      & P    & 0.9591 & \textbf{0.9692} \\
 &                      & R & 0.9592 & \textbf{0.9692} \\
 &                      & F1     & 0.9590 & \textbf{0.9691} \\ \cline{2-5} 
 & \multirow{4}{*}{LPA} & Acc    & 0.9208 & \textbf{0.9625} \\
 &                      & P    & 0.9229 & \textbf{0.9639} \\
 &                      & R & 0.9208 & \textbf{0.9625} \\
 &                      & F1     & 0.9204 & \textbf{0.9624} \\ \hline
\end{tabular}%
}
\end{table}

\begin{table}
\centering
\caption{The prediction results of evolution type by LTSModel and LTSModel (RF only)}
\label{ablation_2}
\resizebox{0.5\textwidth}{!}{%
\begin{tabular}{c|c|ccc}
\hline
Datasets &
  \begin{tabular}[c]{@{}c@{}}Community \\ detection \\ algorithm\end{tabular} &
  Metrics &
  \begin{tabular}[c]{@{}c@{}}LTSModel \\ (RF only)\end{tabular} &
  LTSModel \\ \hline
\multirow{8}{*}{SuperUser}     & \multirow{4}{*}{LMC-APR} & Acc    & 0.9448 & \textbf{0.9692} \\
                               &                          & P    & 0.9447 & \textbf{0.9692} \\
                               &                          & R & 0.9448 & \textbf{0.9692} \\
                               &                          & F1     & 0.9439 & \textbf{0.9691} \\ \cline{2-5} 
                               & \multirow{4}{*}{LPA}     & Acc    & 0.9528 & \textbf{0.9660} \\
                               &                          & P    & 0.9530 & \textbf{0.9661} \\
                               &                          & R & 0.9528 & \textbf{0.9660} \\
                               &                          & F1     & 0.9522 & \textbf{0.9659} \\ \hline
\multirow{8}{*}{StackOverFlow} & \multirow{4}{*}{LMC-APR} & Acc    & 0.9022 & \textbf{0.9130} \\
                               &                          & P    & 0.9043 & \textbf{0.9225} \\
                               &                          & R & 0.9022 & \textbf{0.9130} \\
                               &                          & F1     & 0.8988 & \textbf{0.9112} \\ \cline{2-5} 
                               & \multirow{4}{*}{LPA}     & Acc    & 0.9218 & \textbf{0.9697} \\
                               &                          & P    & 0.9229 & \textbf{0.9702} \\
                               &                          & R & 0.9218 & \textbf{0.9697} \\
                               &                          & F1     & 0.9205 & \textbf{0.9696} \\ \hline
\multirow{8}{*}{SinaHotTopics} & \multirow{4}{*}{LMC-APR} & Acc    & 0.9392 & \textbf{0.9692} \\
                               &                          & P    & 0.9396 & \textbf{0.9692} \\
                               &                          & R & 0.9392 & \textbf{0.9692} \\
                               &                          & F1     & 0.9380 & \textbf{0.9691} \\ \cline{2-5} 
                               & \multirow{4}{*}{LPA}     & Acc    & 0.9575 & \textbf{0.9625} \\
                               &                          & P    & 0.9586 & \textbf{0.9639} \\
                               &                          & R & 0.9575 & \textbf{0.9625} \\
                               &                          & F1     & 0.9566 & \textbf{0.9624} \\ \hline
\end{tabular}%
}
\end{table}

\section{Conclusion}
This paper proposes a two-stage model, LTSModel, for predicting community evolution in interactive networks, thereby forecasting the evolution of collective interactive behaviors because they are reflected by communities. LTSModel distinguishes itself from existing models in two primary ways. Firstly, it unifies the prediction of evolution type using a multi-class classifier and the prediction of evolution extent using a regressor within a single framework, enabling simultaneous predictions of both evolution aspects for a given community. Secondly, LTSModel comprehensively featurizes communities by leveraging community’s characteristics in the interactive network and the positional properties of the community members in the friendship network, enhancing the representativeness of the feature vector. In addition, LTSModel utilizes a hybrid strategy to differentiate between the indistinguishable evolution types \textit{continue} and \textit{expand}, thereby improving predictive performance. Experiments across three datasets demonstrate significant superiority of LTSModel over other models, validating its efficacy in community evolution prediction.

While LTSModel integrates evolution type and extent predictions in one framework, the two components are not jointly optimized during training, which can lead to errors in the evolution extent prediction due to its reliance on the output of the evolution type prediction. Joint optimization of these two components represents a potential avenue for future research to further enhance the predictive performance. Besides, utilizing deep learning techniques to embed communities into low-dimensional vectors for the evolution prediction is an intriguing research direction, given deep learning’s proven success across various domains.

\section*{Acknowledgement}
This work was supported by the Key R\&D Projects of Sichuan Province (No. 2021YFG0333), Natural Science Foundation of China (No. 61802034), and Natural Science Foundation of Sichuan Province (Project No. 2024NSFSC0502). We are grateful for their financial support, which made this study possible.

\bibliographystyle{unsrt}
\bibliography{references}

\begin{thebibliography}{10}

\bibitem{hu2016local}
Yanmei Hu, Bo~Yang, and Chenyang Lv.
\newblock A local dynamic method for tracking communities and their evolution in dynamic networks.
\newblock {\em Knowledge-Based Systems}, 110:176--190, 2016.

\bibitem{takaffoli2014community}
Mansoureh Takaffoli, Reihaneh Rabbany, and Osmar~R Za{\"\i}ane.
\newblock Community evolution prediction in dynamic social networks.
\newblock In {\em 2014 IEEE/ACM International Conference on Advances in Social Networks Analysis and Mining (ASONAM 2014)}, pages 9--16. IEEE, 2014.

\bibitem{que2015scalable}
Xinyu Que, Fabio Checconi, Fabrizio Petrini, and John~A Gunnels.
\newblock Scalable community detection with the louvain algorithm.
\newblock In {\em 2015 IEEE international parallel and distributed processing symposium}, pages 28--37. IEEE, 2015.

\bibitem{makris2020distributed}
Christos Makris, Georgios Pispirigos, and Ioannis~Orestis Rizos.
\newblock A distributed bagging ensemble methodology for community prediction in social networks.
\newblock {\em Information}, 11(4):199, 2020.

\bibitem{wang2022multi}
Zhao Wang, Qingguo Xu, and Weimin Li.
\newblock Multi-layer feature fusion-based community evolution prediction.
\newblock {\em Future Internet}, 14(4):113, 2022.

\bibitem{dakiche2021tailored}
Narimene Dakiche, Fatima Benbouzid-Si~Tayeb, Karima Benatchba, and Yahya Slimani.
\newblock Tailored network splitting for community evolution prediction in dynamic social networks.
\newblock {\em New Generation Computing}, 39(1):303--340, 2021.

\bibitem{saganowski2019analysis}
Stanis{\l}aw Saganowski, Piotr Br{\'o}dka, Micha{\l} Koziarski, and Przemys{\l}aw Kazienko.
\newblock Analysis of group evolution prediction in complex networks.
\newblock {\em PloS one}, 14(10):e0224194, 2019.

\bibitem{chen2022community}
J~Chen, H~Zhao, X~Yang, and et~al.
\newblock Community evolution prediction based on multivariate feature sets and potential structural features.
\newblock {\em Mathematics}, 10(20):3802, 2022.

\bibitem{revelle2021group}
Matt Revelle, Carlotta Domeniconi, and Ben Gelman.
\newblock Group-node attention for community evolution prediction.
\newblock In {\em Proceedings of the 2021 IEEE/ACM International Conference on Advances in Social Networks Analysis and Mining}, pages 176--183, 2021.

\bibitem{dakiche2018sensitive}
Narimene Dakiche, Fatima Benbouzid-Si Tayeb, Yahya Slimani, and Karima Benatchba.
\newblock Sensitive analysis of timeframe type and size impact on community evolution prediction.
\newblock In {\em 2018 IEEE international conference on fuzzy systems (FUZZ-IEEE)}, pages 1--8. IEEE, 2018.

\bibitem{man2019multi}
Junyi Man, Jinrong Zhu, and Langcai Cao.
\newblock Multi-step community evolution prediction methods via marcov chain and classifier chain.
\newblock In {\em 2019 Chinese Control Conference (CCC)}, pages 7950--7955. IEEE, 2019.

\bibitem{boujlaleb2020feature}
Loubna Boujlaleb, Ali Idarrou, and Driss Mammass.
\newblock Feature selection for community evolution prediction in location-based social network: Gowalla and brightkite.
\newblock In {\em Advances in Smart Technologies Applications and Case Studies: Selected Papers from the First International Conference on Smart Information and Communication Technologies, SmartICT 2019, September 26-28, 2019, Saidia, Morocco}, pages 404--412. Springer, 2020.

\bibitem{cheng2018novel}
Jiujun Cheng, Xiao Wu, Mengchu Zhou, Shangce Gao, Zhenhua Huang, and Cong Liu.
\newblock A novel method for detecting new overlapping community in complex evolving networks.
\newblock {\em IEEE Transactions on Systems, Man, and Cybernetics: Systems}, 49(9):1832--1844, 2018.

\bibitem{yuan2022community}
Jiahao Yuan, Zhao Li, Pengcheng Zou, Xuan Gao, Jinwei Pan, Wendi Ji, and Xiaoling Wang.
\newblock Community trend prediction on heterogeneous graph in e-commerce.
\newblock In {\em Proceedings of the Fifteenth ACM International Conference on Web Search and Data Mining}, pages 1319--1327, 2022.

\bibitem{hu2023differential}
Yanmei Hu, Min Lu, Xiangtao Li, and Biao Cai.
\newblock Differential evolution based on network structure for feature selection.
\newblock {\em Information Sciences}, 635:279--297, 2023.

\bibitem{hu2022exhaustive}
Yanmei Hu, Bo~Yang, Bin Duo, and Xing Zhu.
\newblock Exhaustive exploitation of local seeding algorithms for community detection in a unified manner.
\newblock {\em Mathematics}, 10(15):2807, 2022.

\bibitem{lpa2019}
Sara~E. Garza and Satu~Elisa Schaeffer.
\newblock Community detection with the label propagation algorithm: A survey.
\newblock {\em Physica A: Statistical Mechanics and its Applications}, 534(122058), 2019.

\bibitem{amerini2019social}
Irene Amerini, Chang-Tsun Li, and Roberto Caldelli.
\newblock Social network identification through image classification with cnn.
\newblock {\em IEEE access}, 7:35264--35273, 2019.

\bibitem{hancock2020catboost}
John~T Hancock and Taghi~M Khoshgoftaar.
\newblock Catboost for big data: an interdisciplinary review.
\newblock {\em Journal of big data}, 7(1):94, 2020.

\bibitem{ramraj2016experimenting}
Santhanam Ramraj, Nishant Uzir, R~Sunil, and Shatadeep Banerjee.
\newblock Experimenting xgboost algorithm for prediction and classification of different datasets.
\newblock {\em International Journal of Control Theory and Applications}, 9(40):651--662, 2016.

\bibitem{ke2017lightgbm}
Guolin Ke, Qi~Meng, Thomas Finley, Taifeng Wang, Wei Chen, Weidong Ma, Qiwei Ye, and Tie-Yan Liu.
\newblock Lightgbm: A highly efficient gradient boosting decision tree.
\newblock {\em Advances in neural information processing systems}, 30, 2017.

\bibitem{parbat2020python}
Debanjan Parbat and Monisha Chakraborty.
\newblock A python based support vector regression model for prediction of covid19 cases in india.
\newblock {\em Chaos, Solitons \& Fractals}, 138:109942, 2020.

\end{thebibliography}
\end{document}